\documentclass{ws-procs9x6}

\begin{document}

\title{AGN Observations with the MAGIC Telescope}

\author{C. Bigongiari$^*$ for the MAGIC Collaboration}

\address{Universit\`a di Padova and INFN Padova, \\
Via F.Marzolo 8 - I35131 Padova, Italy \\
$^*$E-mail: ciro.bigongiari@pd.infn.it \\
wwwmagic.mppmu.mpg.de}

\begin{abstract}

MAGIC is presently the imaging atmospheric Cherenkov telescope
with the largest reflecting surface and the lowest energy threshold.
MAGIC concluded its first year of regular observation in April 2006. 
During this period and the preceding commissioning phase, 
25 Active Galactic Nuclei have been observed and VHE $\gamma$-ray emission 
has been confirmed by 4 of them. Two more AGNs have been detected as $\gamma$-ray sources  
with high statistical significance for the first time.
We report in this paper the results obtained analyzing data of the detected sources.
Temporal and spectral properties of detected signals are shown and discussed.

\end{abstract}


\bodymatter

\section{Introduction}

MAGIC (Major Atmospheric Gamma Imaging Cherenkov) telescope \cite{Baixeras_2004,Cortina_2005}, 
located on the Canary Island La Palma (2200 m a.s.l., $28.4^{\prime}$N, $17.54^{\prime}$W), 
is currently the largest imaging air Cherenkov telescope in operation. 
The MAGIC construction was completed in Fall 2003 and after a commissioning phase of about one year
MAGIC started its first regular observation cycle in April 2005. 
According to detailed simulation of atmospheric showers and detector response the trigger threshold is around 60~GeV
for low zenith angle observations \cite{Majumdar_2005} while 
the analysis threshold is about $E_{Th} = 100$~GeV in the same conditions.
MAGIC integral flux sensitivity has been calculated from Monte Carlo simulation and results in 
about 5\% of ${\Phi}_{Crab}$ at $E > 100$~GeV and 2\% of ${\Phi}_{Crab}$ at $E > 1$~TeV
\cite{Majumdar_2005}. 
The angular resolution has been estimated applying the DISP method to Crab data
and results in about $0.1^{\circ}$ for $\gamma$-ray events with $E > 200$~GeV  \cite{Domingo_2005}.
The accuracy in the determination of the point-source position improves as the square root of the number of collected events 
and is ultimately limited by tracking accuracy ($\simeq 0.02^{\circ}$)
\cite{Riegel_2005}.
The energy resolution has been estimated from Monte Carlo data and results in ${\Delta}E/E \simeq 30\%$ at $E = 100$~GeV and 
${\Delta}E/E \simeq 20\%$ for $E > 1$~TeV 
\cite{Wagner_2005}.  
The systematic errors on the measured flux were
estimated to be around 50\% for the absolute flux level and
0.2 for the spectral index.

\section{Observed Blazars} 

MAGIC, during its first year of regular data taking, observed a sample 
of Blazars, mainly HBLs, at redshifts $z < 0.3$. This sample was 
chosen selecting northern Blazars with the highest 
expected VHE (here defined as $E > 100 \mathrm{GeV}$) fluxes according to leptonic 
and hadronic models of $\gamma$-ray emission.
Moreover, considering possible correlations between VHE emission 
and optical/X-ray emission, 
MAGIC performed Target of Opportunity
observations whenever alerted by optical/X-ray telescopes.
We present here the spectral properties and the temporal behavior 
of the detected sources:  Mkn~421, Mkn~501, Mkn~180, 1ES1959+650, 1ES1218+304
and PG1553+113.

\subsection{Markarian 421}

Mkn~421 (redshift $z = 0.030$) was the first extragalactic VHE
$\gamma$-ray source detected by Whipple \cite{Whipple_Mkn421} in 1992.
Many observations of this source were performed since then showing 
flux variations larger than one order of magnitude and flares with 
doubling times as short as 15 minutes \cite{Gaidos_Mkn421_Flares}.  
A significant correlation between X-ray and $\gamma$-ray flux 
has been detected during multi wavelength campaigns involving 
Cherenkov telescopes and X-ray detectors \cite{Krawczynski_Mkn421}.  
\begin{figure}
\centering
  \includegraphics[width=0.9\textwidth]{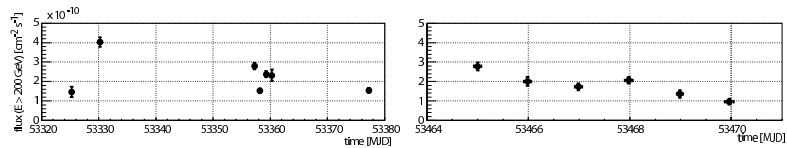}
\caption{ Light curve for Mkn~421 from November 2004
to April 2005. Each data point is the nightly averaged integral flux above 200~GeV. Left panel: data from November 2004 to January 2005.  Right
panel: data for April 2005. } 
\label{fig:Mkn421_LightCurve} 
\end{figure}

MAGIC observed Mkn~421 for 19 nights and an overall observation time of 15.5 hours. 
All the data were taken at zenith angle below ($30^\circ$) with the only exception of 1.5
hours taken in December 2005 at $42^\circ < ZA < 55^\circ$ during
simultaneous observations with H.E.S.S. \cite{HESS_MAGIC_Mkn421}.  

\begin{figure}[thb]
\begin{minipage}[t]{0.46\textwidth}
	\centering 
  	\includegraphics[width=0.9\textwidth,angle=0,clip]{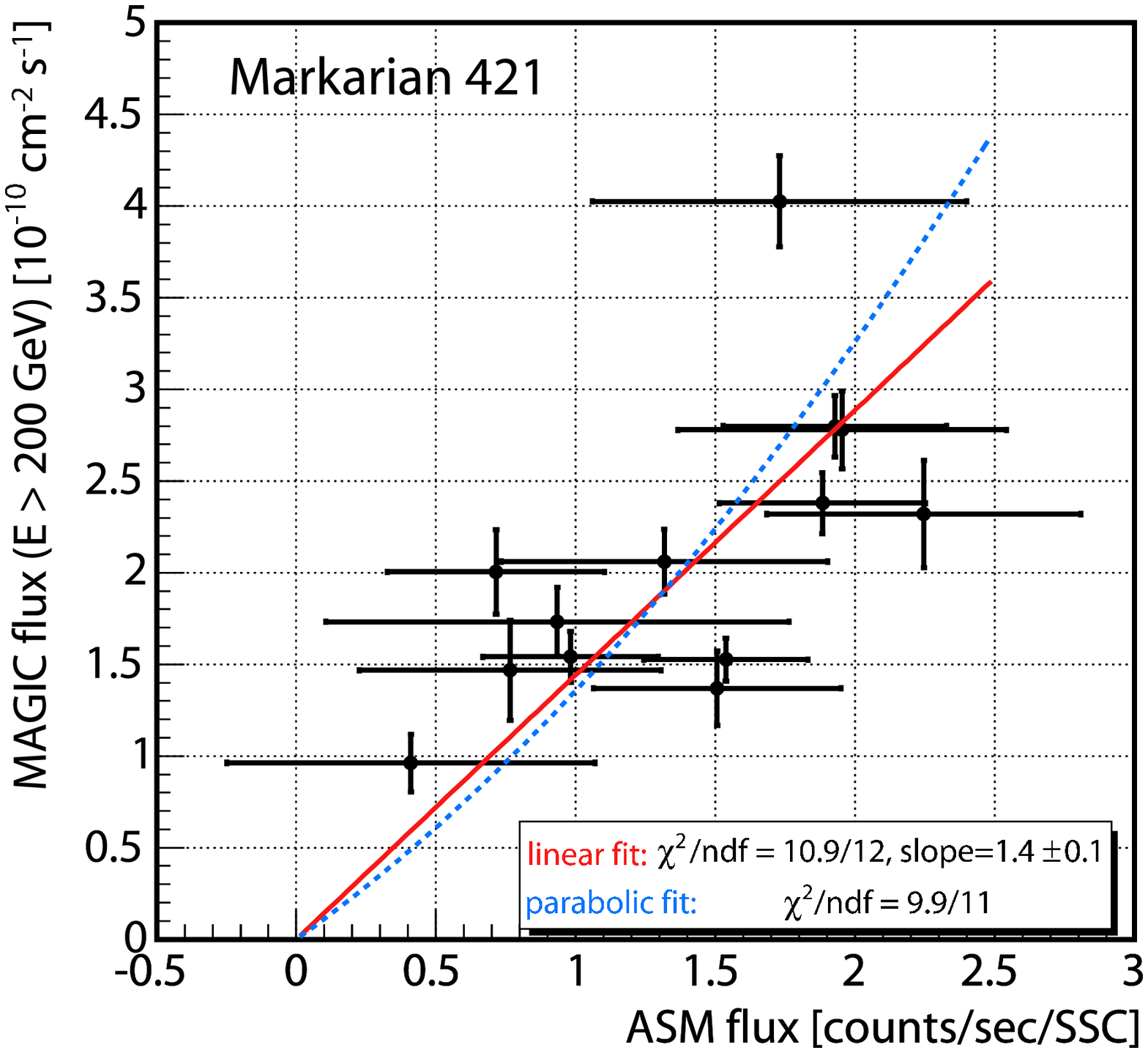}
			\caption{
							Correlation plot between VHE $\gamma$-ray flux above 200~GeV measured by MAGIC and  
    					and X-ray counts by ASM-RXTE for Mkn~421.}
			\label{fig:Mkn421_Xray}   
\end{minipage}
\ \hspace{5mm} \
\begin{minipage}[t]{0.46\textwidth}
\centering
  \includegraphics[width=0.95\textwidth,angle=0,clip]{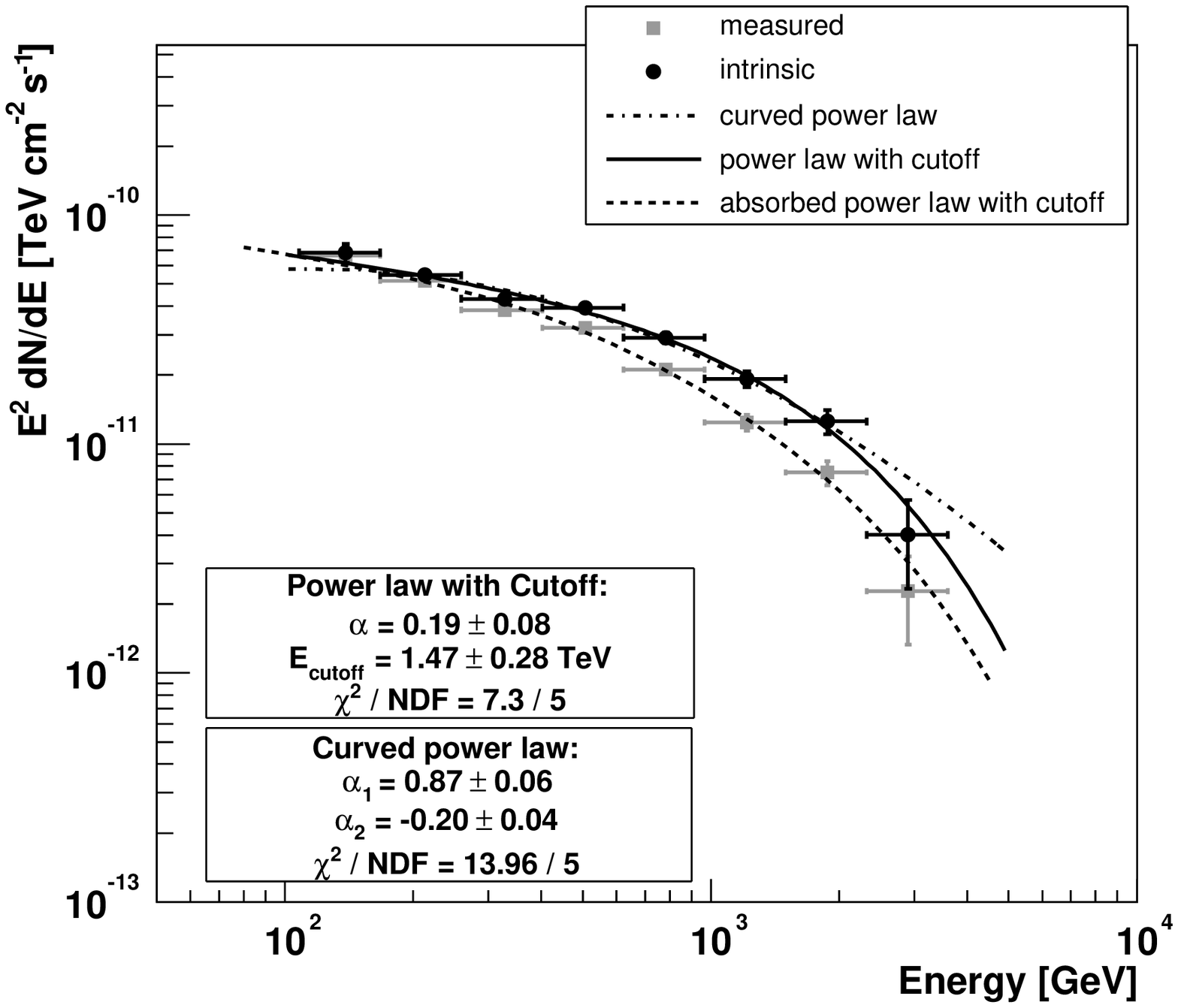}
	\caption{
         	Measured (gray squares) and de-absorbed (black circles) energy flux 
         	of Mkn~421. 
 %        	Solid line: fit (1) to the intrinsic spectrum using a power 
 %        	law with a cut-off. 
 %        	Dashed-dotted line: fit (2) to the intrinsic spectrum using 
 %        	a curved power law. A dashed line indicates the expected absorbed 
 %        	spectrum using the result of fit (1). 
 %        	Fit parameters of the intrinsic spectrum 
 %        	are shown in the inlays.
 					}
	\label{fig:Mkn421_Spectra}    
\end{minipage}
\end{figure}
%%%%%%%%%%%%%%%%%%%%%%%%%%%%%%%%%%%%%%%%%%%%%%%%%%%%%%%%%%%%%%%%%%%%%%%%%%%%%%%%%%%%%%%%%55

During MAGIC observations Mkn~421 flux above 200~GeV ranged from 0.5 to 2 Crab units (see
Fig.~\ref{fig:Mkn421_LightCurve}). Significant flux variations up to a factor four overall
and up to a factor two between successive nights can be seen.
A clear correlation between the X-ray flux, measured by the
All-Sky-Monitor on-board the RXTE satellite, and the VHE $\gamma$-ray flux measured by MAGIC 
can be seen in Fig.~\ref{fig:Mkn421_Xray}.  
%Both a straight line and a parabola (forced to go through the axes origin), can fit the data.
The energy density distribution of gammas from Mkn~421, 
that is the differential photon spectra multiplied by $E^2$, 
is shown in Fig.~\ref{fig:Mkn421_Spectra} both for the measured spectrum 
and the de-absorbed one 
(i.e. corrected for the effect of extragalactic absorption). 
The de-absorbed spectrum is curved, clearly indicating that the curvature in the measured
spectrum is not caused by the absorption of the VHE $\gamma$-rays by the EBL photons but has an intrinsic origin.  
%A fit with a pure
%power law with a exponential cut-off as well as a fit with a curved power law
%indicate a flattening of the spectrum towards 100~GeV.  
For further details about Mkn~421 data analysis see \cite{MAGIC_Mkn421}.

\subsection{Markarian 501} 

\begin{figure}[thb]
\centering
  \includegraphics[width=0.9\textwidth,angle=0,clip]{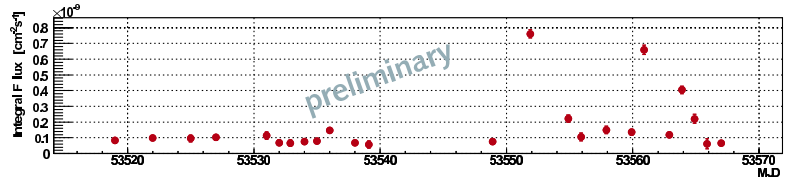}
\caption{Light curve for Mkn~501 from June to July 2005. Each data point is the night averaged
 integral flux above 200~GeV.}
\label{fig:Mkn501_LightCurve}
\end{figure}

Mkn~501 (redshift $z = 0.034$) was the second extragalactic VHE $\gamma$-ray source 
discovered by Whipple \cite{Whipple_Mkn501} in 1996.
Following observations in 1997 showed that 
Mkn~501 integral flux above 1~TeV can 
reach 10 Crab units \cite{HEGRA_Mkn501}
and can drastically change on timescales of 0.5 days. 

\begin{figure}[htb]
	\begin{minipage}[t]{0.46\textwidth}
		\centering
  	\includegraphics[width=0.9\textwidth,angle=0,clip]{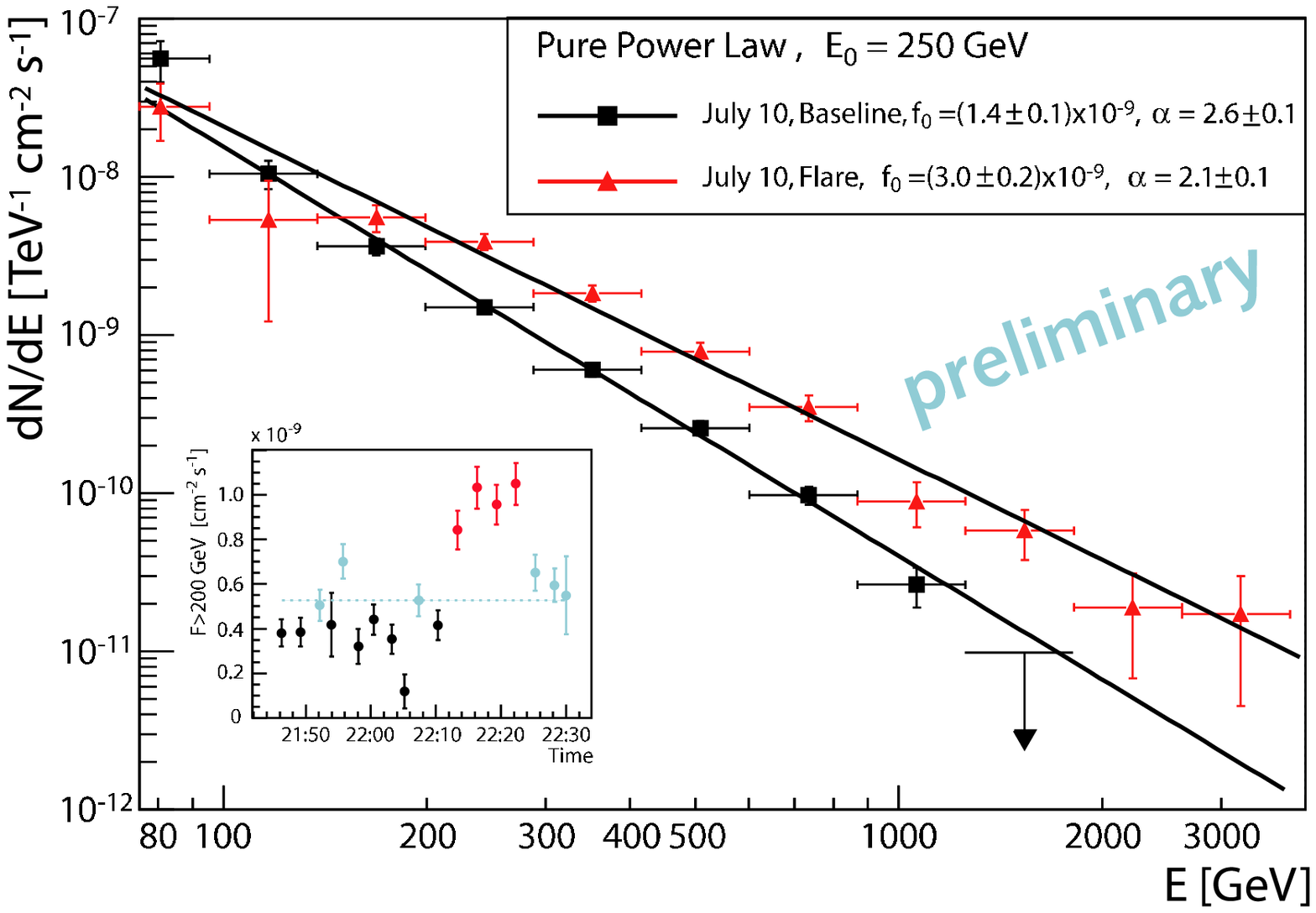}
		\caption{Differential energy distribution of Mkn~501 events
		recorded on 10th July 2005. 
		The rapid variation of the flux level and corresponding change in the shape 
		of the energy spectrum is clearly visible.}
		\label{fig:Mkn501_Flare_Spectra} 
	\end{minipage}
\ \hspace{5mm} \	
	\begin{minipage}[t]{0.46\textwidth}	
	 \centering
  	\includegraphics[width=0.95\textwidth,angle=0,clip]{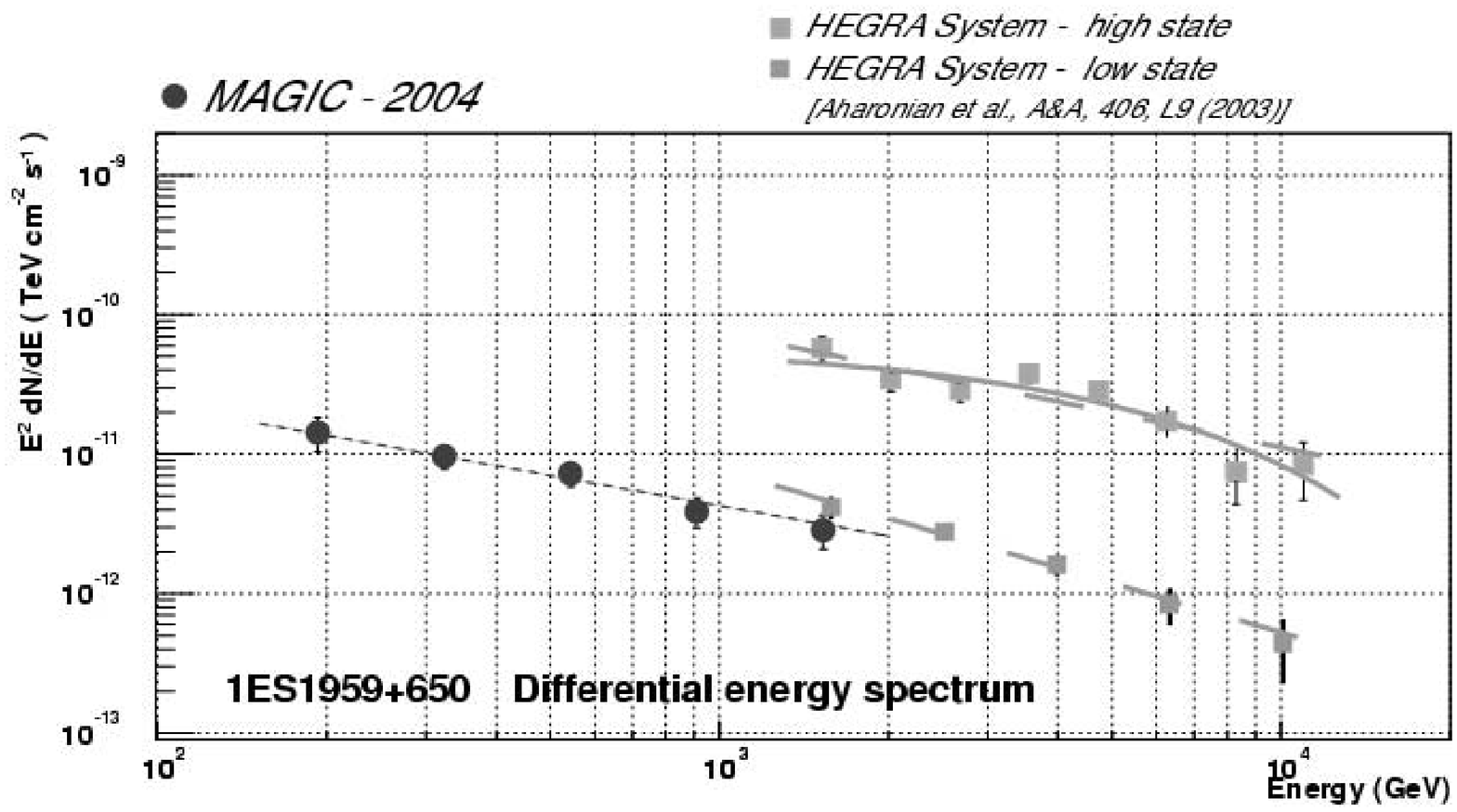}
		\caption{
		Differential energy spectrum of 1ES1959+650. For comparison, 
		the differential energy spectra measured by HEGRA when the source was 
		in two different activity levels are also shown.
		}
	 \label{fig:1ES1959_Spectra} 
	\end{minipage}
\end{figure}

MAGIC observed Mkn~501 between June and July 2005 for a total of 24 nights and 32.2 hours of
observation time. Mkn~501 was in a rather low flux state (integral flux above 200~GeV around 0.4 Crab units) 
when MAGIC started its observation. Suddenly, on 30th Jun, its flux reached 4
Crab units, see  Fig~\ref{fig:Mkn501_LightCurve}.
This flare stimulated further observations which were performed in the following days 
also in the presence of moonshine to extend the time coverage. 
The source was found in high state (integral flux above 2 Crab units) on two more nights.
In particular a flare with doubling time as short as 5 minutes or less was detected on the
night of 10th July 2005, see the inlay in Fig.~\ref{fig:Mkn501_Flare_Spectra}. 
The high source flux together with the MAGIC high sensitivity allowed the measurement 
of the source spectrum in time intervals as short as 10 minutes. 
A significant hardening of the spectrum as the flux grows was found, 
confirming previous indications by Whipple \cite{Whipple_Mkn501}.
It is worth noticing that this is the first time 
that spectral hardening has been measured on time scales of $\sim 10$ minutes.  
A detailed publication on the
results of the Mkn~501 data analysis is in preparation.

\subsection{Markarian 180}

Mkn~180 (1ES1133+704) at redshift $z = 0.045$ is the second extragalactic $\gamma$-ray source discovered by MAGIC. 
Previously this source had been observed by HEGRA and Whipple collaborations but these observations 
resulted only in upper limits on the VHE $\gamma$-ray flux \cite{Aharonian_Mkn180,Horan_Mkn180}.
MAGIC observed Mkn~180 for 9 nights after an alert received by the KVA
telescope, on 23rd March 23 2006.
In the following nights MAGIC observed this source for 14.4 hours.
Only 11.1 hours survived to the quality cuts applied to remove runs with unusual trigger rate, usually related to  
not optimal atmospheric conditions.  
This sample was enough to find a clear signal at $6.5 \sigma$ level with 271 excess events and to measure  
the integral flux above 200~GeV, $\Phi(E>200 \mathrm{GeV}) = (2.3 \pm 0.7) \times 10^{-7} m^{-2} s^{-1}$, which corresponds to 10\% of the Crab Nebula flux as measured by MAGIC. 
No evidence of flux variation was found on a daily scale. 
\begin{figure}[thb]
		\centering 
			\includegraphics[width=0.8\textwidth,angle=0,clip]{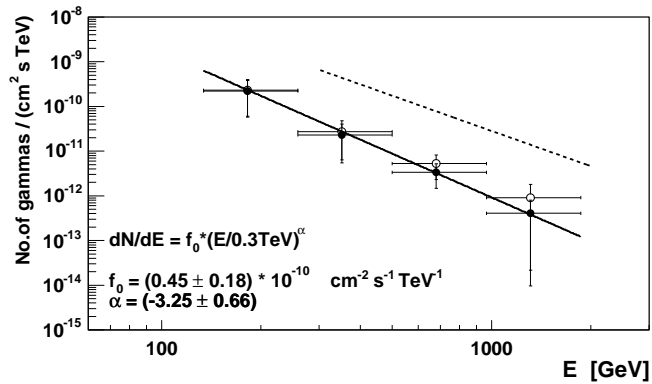}
				\caption{Differential energy spectrum of Mkn~180, measured by MAGIC (full circles) and 
				corrected for the effect of extragalactic absorption (open circles).
				The black line represents a power-law fit to the measured spectrum. The fit paremeters are listed in the 
				figure. For comparison, the Crab Nebula spectrum measured by MAGIC is also shown (dotted line).}
		\label{fig:Mkn180_Spectra}
\end{figure}
The differential energy distribution, shown in figure~\ref{fig:Mkn180_Spectra} with filled points, can be fitted 
by a power law with spectral index $\alpha = 4.2 \pm 0.7$. In the same figure also the de-absorbed spectrum 
(i.e. corrected for the effect of the extragalactic absorption)
is shown (open circles). 
%A fit with a simple power law to the de-absorbed spectrum gives a slope with $\alpha = 2.8 \pm 0.7$.
For further details on this analysis see \cite{MAGIC_Mkn180}.

\subsection{1ES1959+650}

The first hint of VHE $\gamma$-ray emission from 1ES1959+650 ($z = 0.047$) was claimed by 
the Seven Telescope Array collaboration in 1998 \cite{Seven_Telescope_Array_1ES1959}.  
This claim was later confirmed both by Whipple and HEGRA collaborations 
\cite{Whipple_1ES1959,HEGRA_1ES1959} which observed this source in May 2002 
when its X-ray flux was much higher than the usual level. 
During observations in 2002 a so-called {\it orphan flare}, 
i.e. a VHE $\gamma$-ray activity in the absence of high activity in X-rays, 
was observed.  
Orphan flares are very interesting because they could be 
an indication of hadronic acceleration in Blazars. They  
are not expected, in fact, within the SSC model. 
MAGIC observed 1ES1959+650 when it was still 
in the commissioning phase, therefore not at its best performances.  
Nevertheless a clear signal at $8.2\sigma$ significance level was found
and the energy spectrum was measured down to 180~GeV for the first time.
The energy spectrum between 180~GeV and 2~TeV can be well fitted 
with a power law with a photon index
$\alpha=\mbox{2.72}\pm\mbox{0.14}$ and is consistent with the one 
measured by HEGRA \cite{HEGRA_1ES1959}, see Fig~\ref{fig:1ES1959_Spectra}.
%\begin{figure}[htb]
%	\centering
%  	\includegraphics[width=0.9\textwidth,angle=0,clip]{1ES1959_Spectra.eps}
%		\caption{
%		Differential energy spectrum of 1ES1959+650. For comparison 
%		the differential energy spectra measured by HEGRA when the source was 
%		in two different activity levels are shown too.
%		}
%	\label{fig:1ES1959_Spectra} 
%\end{figure}
The integral VHE
$\gamma$-ray flux above 180~GeV resulted in 
$(3.73 \pm 0.41) \times 10^{-7} m^{-2}s^{-1}$, 
in agreement with the low state flux measured by HEGRA.
For further details of the analysis of these data and its results see
\cite{MAGIC_1ES1959}.

\subsection{1ES1218+304}

The Blazar 1ES1218+304 at redshift $z = 0.182$ is the first extragalactic VHE $\gamma$-ray source discovered by MAGIC. 
This source has been observed by Whipple since the discovery of Mkn~501 but these observations 
leaded only to an upper limit on the VHE flux \cite{Whipple_1ES1218}. 
HEGRA collaboration also observed this source, but again no detection was claimed and 
only an upper limit on the $\gamma$-ray flux was published \cite{HEGRA_1ES1218}. 
\begin{figure}
	\begin{minipage}[t]{0.46\textwidth}
		\begin{center}
			\includegraphics[width=0.9\textwidth,angle=0,clip]{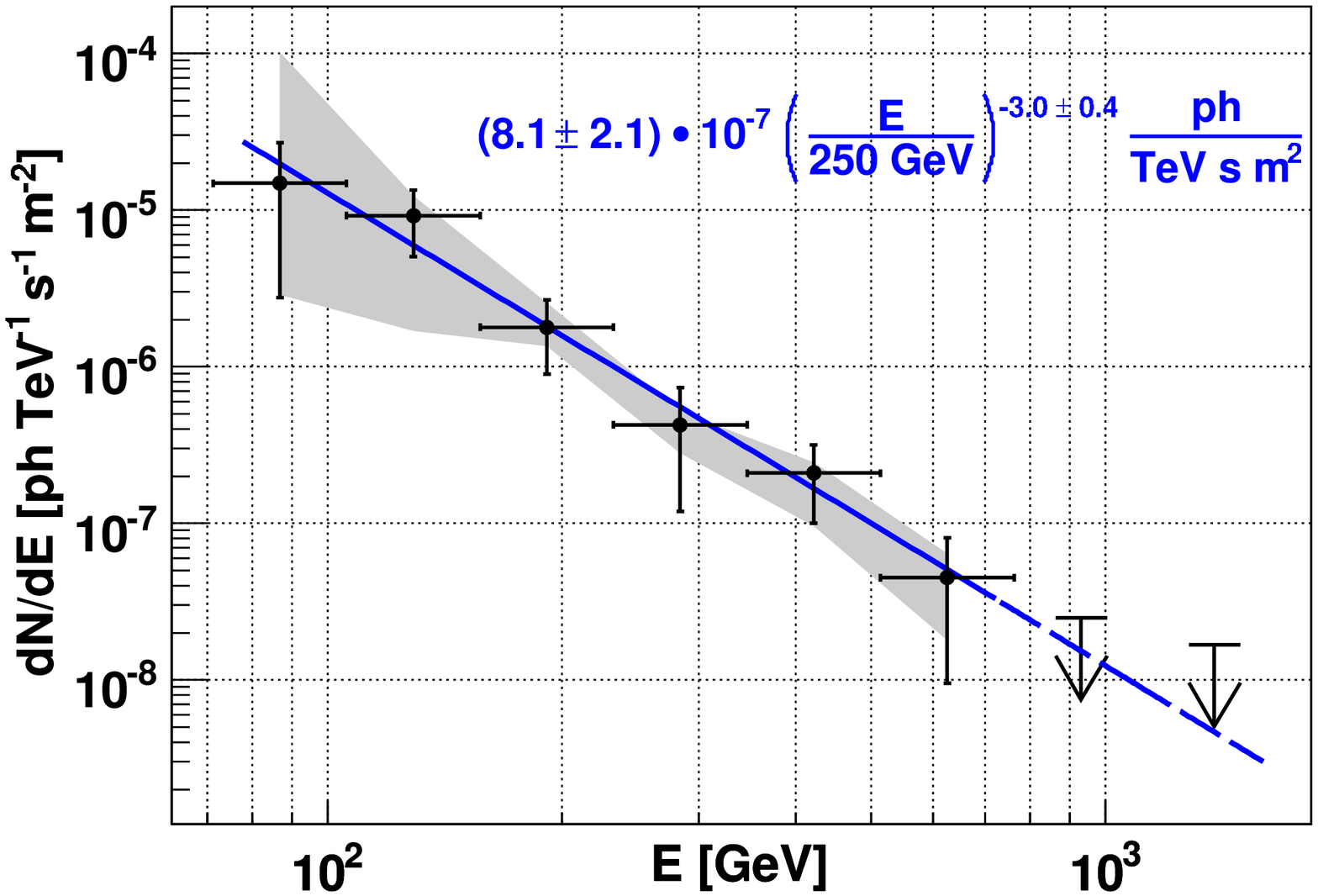}
			\caption{Differential energy spectrum of 1ES1218+304. The upper limits correspond to 90\% confidence level. The 
			grey-shaded region shows the systematic error due to initial MC spectrum analysis cuts.}
			\label{fig:1ES1218_Spectra}
		\end{center}
	\end{minipage}	
\ \hspace{5mm} \
	\begin{minipage}[t]{0.46\textwidth}
		\begin{center}
			\includegraphics[width=0.9\textwidth,angle=0,clip]{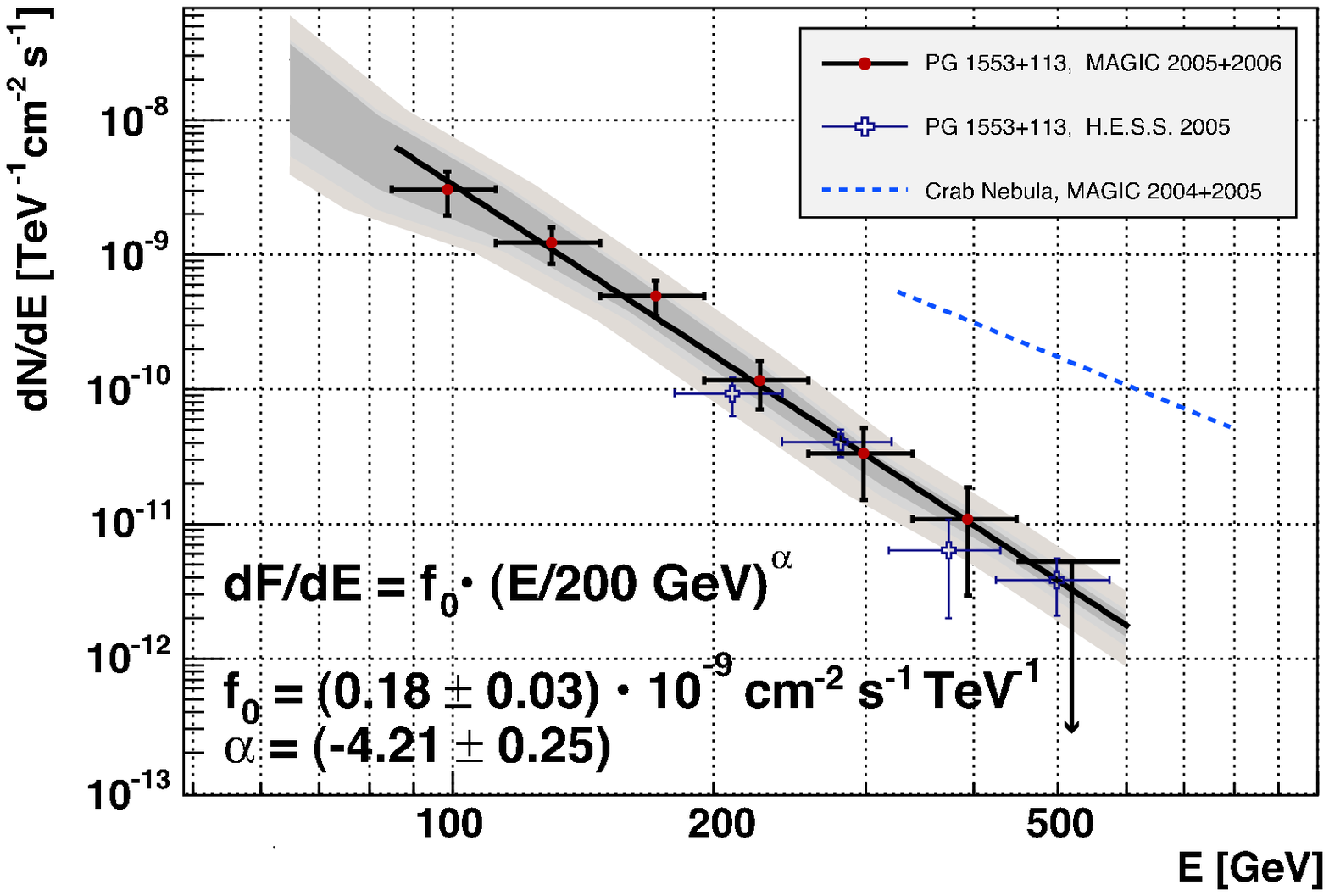}
			\caption{Differential energy spectrum of PG1553+113 as derived from the combined 2005 and 2006 data. The 
			grey-shaded region shows the systematic error due to initial MC spectrum and analysis cuts.}
			\label{fig:PG1553_Spectra}
		\end{center}	
  \end{minipage}
\end{figure}

MAGIC observed 1ES1218+304 for 8.2~hours during seven nights in January 2005.  
An excess of 560 events with statistical
significance of $6.4 \sigma$ was found. 
The night-by-night $\gamma$-ray light curve didn't show any statistically 
significant variations.  
The energy spectrum between 80~GeV and 600~GeV can be well fitted by a power
law with photon index $\alpha=3.0\pm0.4$. 
The integral flux above 100~GeV is $\Phi(E > 100 \mathrm{GeV}) = (8.7 \pm 1.4) \times 10^{7} m^{-2}s^{-1}$
and is below the upper limits at higher energies determined in the past.  
For details about these data analysis see \cite{MAGIC_1ES1218}.

\subsection{PG1553+113}

The first evidence of VHE $\gamma$-ray emission from PG1553+113 (redshift unknown, $z > 0.09$)
was claimed in 2005 by H.E.S.S. collaboration showing an excess at $4 \sigma$ level 
\cite{HESS_1553}. 
MAGIC started to observe this source in 2005 and,
motivated by 
a strong hint of signal in the 2005 data,
continued in 2006
for an overall observation time of 18.8~hours.  
A very clear signal  was detected with a significance of $8.8~\sigma$.  
There is no evidence of $\gamma$-rays 
short term variability, but a factor of three change in the
flux level from 2005 to 2006 was found.  The combined 2005 and 2006
differential energy spectrum for PG1553+113 is well described by a pure power
law with a photon index $\alpha=4.2\pm0.4$, in good agreement with H.E.S.S.
result 
in the overlapping energy range.
For details of the analysis and results see \cite{MAGIC_1553}.

%The measured energy spectrum of Mkn 180 is
%shown in Fig. 6. A fit by a power law gives a photon index
%a = 3.3±0.7. The observed integral flux above 200 GeV is
%11% of the Crab Nebula flux. The attenuation of the spectrum
%caused by the EBL was determined by numerical integration
%of Eq. 2 in [35]. The de-absorbed energy spectrum of
%Mkn 180 is also shown in Fig. 6 (open circles). 

\section{Conclusions and Outlook} 

An overview of the extragalactic VHE $\gamma$-ray sources detected
by MAGIC until May 2006 is given. 
In this period MAGIC detected VHE $\gamma$-ray emission by 6 extragalactic 
sources: Mkn~421, Mkn~501, Mkn~180, 1ES1959+650, 1ES1218+304 and PG1553+113. 
Two of them, 1ES1218+304 and Mkn~180, have been discovered by MAGIC while PG~1553+113 
has been confirmed as $\gamma$-ray source at high significance level 
after the first hint of signal by H.E.S.S.
The energy spectrum of Mkn~421 has been measured down to 100~GeV for the first time
showing a de-absorbed spectrum which clearly flattens toward 100~GeV. 
Mkn~501 has been caught in flaring states in 3 nights with flux higher than 2 Crab units.    
Flux doubling time down to 5 minutes has been observed and spectral hardening 
as the flux increases has been measured on 10 minutes time scale.
Gamma-ray emission from Mkn~180 has been discovered during an optical outburst indicating 
a possible correlation between $\gamma$-ray and optical emission. 
%but no conclusions can be drawn until this source will be observed in low optical state. 
The energy spectrum of 1ES1959+650 has been measured down to 180~GeV, well below previous 
measurements by HEGRA, when the source was in a low activity state.  

Re-observations of the presented sources as
well as the analyses of further observed Blazars are ongoing.

Collaborations with observatories in other spectral ranges have proved to be 
very fruitful and will be extended and strengthened in the near future.

\section{Acknowledgments}
 
We would like to thank the IAC for the excellent
working conditions at the ORM in La Palma. The support of the
German BMBF and MPG, the Italian INFN, the Spanish CICYT, the ETH
research grant TH~34/04~3, and the Polish MNiI grant 1P03D01028 is
gratefully acknowledged. 

%\bibliographystyle{ws-procs9x6}
%\bibliography{ws-pro-sample}
% BibTeX users please use
%\bibliographystyle{spmpsci}


\end{document}